\documentclass[11pt,dvips]{article} 
\usepackage{epsfig,times} 
\usepackage[english]{babel} 

\begin{document} 
\begin{center} 
{\bf \Large  
Feasibility of the correlation curves method in calorimeters of different types 
}  
\vspace{2mm} 

{\it \large E.A. Grushevskaya, I.A. Lebedev, A.I. Fedosimova} 

\vspace{2mm} 

{\large Institute of Physics and Technology, Almaty, Kazakhstan} 

\vspace{2mm} 

\end{center} 

{\small  
The simulation of the development of cascade processes in calorimeters of different types  for the implementation of energy measurement  
by correlation curves method, is carried out. Heterogeneous calorimeter has a significant transient effects, associated with the difference of the critical 
energy in the absorber and the detector. The best option is a mixed calorimeter, which has a target block, leading to the rapid development 
of the cascade, and homogeneous measuring unit. Uncertainties of energy reconstruction by presented mixed calorimeter on the base of  the 
correlation curves methodology, is less than 10 percent. 
} 

\large

\section{Introduction} 

Researches of the characteristics of cosmic rays (CR) on the basis of "direct" measurements outside the atmosphere on the spacecrafts or high-altitude balloons can allow to solve many problems of particle physics, cosmic ray physics and astrophysics. Consequently, this region has a significant interest and dynamism to the development of theoretical and experimental studies worldwide \cite{1}-\cite{5}.

Direct experiments on high-altitude balloons: JACEE and RUNJOB, which were aimed to studying the elemental composition of cosmic rays, gave very similar spectra of protons, but very different spectra of helium nuclei \cite{6}. Data on heavier nuclei have weak statistics due to the relatively high energy threshold of the applied methodology. Great hopes were pinned on experiments ATIC \cite{7} and CREAM \cite{8}, however, contradictory data were obtained from the spectral indices of the basic elements of cosmic rays, which do not provide a consistent picture of the processes occurring in the sources of cosmic rays and their propagation to the Earth.

The main advantage of direct experiments is the ability to measure the charge of the incident particle. It is much harder to use outside the atmosphere a energy detector for particles with energies of $E >10^{12}$ eV. 

For particles at relatively low energies $E<10^{12}$ eV, it can use different techniques such as the technique of threshold Cherenkov 
counters, the method of the RICh detector, the technique of magnetic mass spectrometer. At higher energies, principal threshold effects , 
limiting their application, are arised.

The best option, for energy measurements of different nuclei in a wide energy range ($E >10^{12}$ eV) at present, remains only the method of ionization calorimeter \cite{9}.

The main problem with this method of energy measurement is massive installation because the calorimeter must have a sufficiently depth to build the cascade curve. This greatly complicates the possibility of using such a device in the space industry.

In this regard, a more promising  approach, for the determination of the energy on the
 basis of direct measurements of CR, is the use of thin calorimeter. In a thin calorimeter the entire cascade of secondary particles is not fixed, and it is recorded only the beginning. The energy is determined on the basis of the analysis of the size of the cascade, because the number of particles in the cascade is almost proportional to the energy of the primary particle \cite{10}.

In the detector of space experiment NUCLEON \cite{11} the weight reduction of the calorimeter is achieved by using kinematic methods for 
determining the energy of the primary particle. Calculations and test experiments at the accelerator showed that the accuracy of 
the determination of the energy will be about 50 percent.

This low accuracy is due to the fact that the results of the energy measurement is particularly sensitive to fluctuations in the cascade 
development process and they depend significantly on the mass of the primary nucleus.

In \cite{12} the technique of correlation curves is developed. The approach allows to significantly increase the accuracy of the energy measurement of high-energy nuclei, by analyzing the internal correlations of the development of the cascade, which significantly reduces the effect of fluctuations in the development stage to the results of measurement of energy. 

The approach is based on using the correlation curves, which connect the logarithm of the number of particles at a certain level observations $log N$ and the difference of the number of particles at two observation levels, separated by a layer of absorber $dN=logN_1-logN_2$.

The main idea of the approach was based on the assumption about the universality of the development of the cascade. 
This assumption was based on the following. Among the various causes of fluctuations, the greatest effect is achieved from 
fluctuations of the first act of interaction, because in subsequent interactions, the number of participating particles is large, 
and the fluctuations of individual interactions more or less compensate each other. Therefore, all the cascades, formed by the primary 
particles of the same mass and energy, starting from a certain moment, develop almost the same. This assumption was confirmed by the 
simulation results.

In this paper, some technical issues of the implementation of the methodology of the correlation curves for thin calorimeter, is considered. 

Simulations is performed on the basis of the software package GEANT4. Throughout the development of the cascade it was “detected” the number of particles, their spatial coordinates, energy and time. As various options the heterogeneous, homogeneous and mixed patterns were considered.  

\section{ Heterogeneous calorimeter with a thin detector}

The simulated heterogeneous calorimeter has a multi-layer structure. Each layer consists of 5 mm of lead $Pb$ and 5 mm of silicon $Si$. In this calorimeter, $Pb$ plays the role of dense matter, which is developing the cascade. Silicon is used only as a detector. To understand the dynamics of the cascade process in the first stage it was considered the calorimeter, consisting of 50 layers - $50x(0.5 Pb+0.5 Si)$.

Lead is one of the most commonly used material because it has a low amount of radiation units $t$=6.4 g/cm$^2$. Taking into account the high density of lead, which is 11.34 g/cm$^3$, the thickness of the absorber of  0.56 cm corresponds to one radiation unit. It is very low value, which allow to get the thin calorimeter. Unfortunately, $Pb$ may not perform as a detector and on its basis it is impossible to build a homogeneous calorimeter.

The most commonly used material for the detector is the silicon. The density of silicon is 2.35 g/cm$^3$, and the amount of radiation units is 22.2 g/cm$^2$. Thus, for the passage of the absorber in one radiation unit, it must use the silicon thickness of 9.4 cm. 

In experiment, the development of the cascade in the lead is not fixed. And it is possible to observe  the development of the cascade only in the detector. Modeling gives such an opportunity. This is a significant plus.

Taking into account the possibilities of the modeling, the measurements of characteristics of particles in the detector and absorber, was carried out. The main development of the cascading process takes place in the lead (each layer approximately corresponds to one radiation unit), and the detector is intended only for determination of the number of particles in the cascade, as its depth is only 0.05 t.

The cascade development process in the calorimeter is presented in Figure 1. 

\begin{figure}[h] 
\begin{center} 
\includegraphics*[width=0.6\textwidth,angle=0,clip]{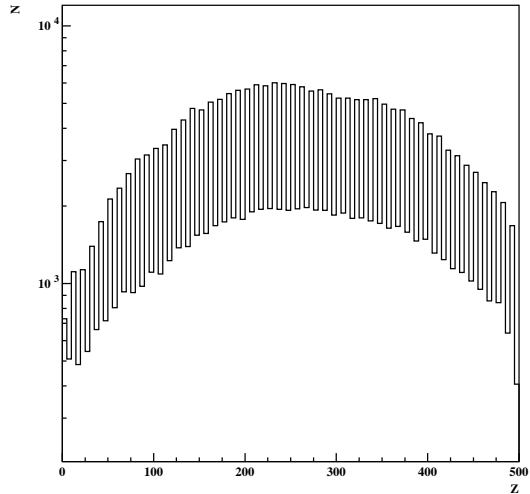} 
\caption{\label{fig1}  
The cascade curve, formed by $10^{12}$ eV iron nuclei in the calorimeter of $50x(0.5Pb+0.5Si)$.} 
\end{center} 
\end{figure} 

As it can be seen from Figure 1, the cascade curves are dramatically changing structure with huge oscillations in the number of particles 
in the absorber and in the detector. And as a consequence, this calorimeter has significant errors at energy reconstruction and 
significantly influenced of the hardware on the measurement results due to the low thickness of the measuring layer.

This is due to the significant difference between value of the critical energy in the material of the absorber and detector. 
The main ionization in the calorimeter is generated by relativistic electrons from the developing cascade. The number of electrons 
in the maximum of the cascade depends on the ratio of the energy of the primary particles and the critical energy in the substance. 
The critical energy in the lead and in silicon differs significantly. In the lead it is 7.4 MeV, and in silicon it is 37.5 MeV. Thus, 
the number of particles in the lead and in silicon differs in 5 times. 

Thus, in passing from lead into the silicon the particles are quickly absorbed and the balance is disturbed.

\section{ Heterogeneous calorimeter with a thick detector}
	
In the second stage the heterogeneous calorimeter with increased detector layer, was considered. Simulated calorimeter has the 
multilayer structure. Each layer includes  1 cm of lead and 19 cm of silicon. That is, in t-units the detector layer roughly 
corresponds to the layer of the absorber. In this case after the transition from lead into silicon the registration of particles can 
be performed after the relaxation process of the absorption of the particles. The depth of the detector gives such an opportunity.

The development of cascade process in the calorimeter is presented in Figure 2.

\begin{figure}[tbh] 
\begin{center} 
\includegraphics*[width=0.6\textwidth,angle=0,clip]{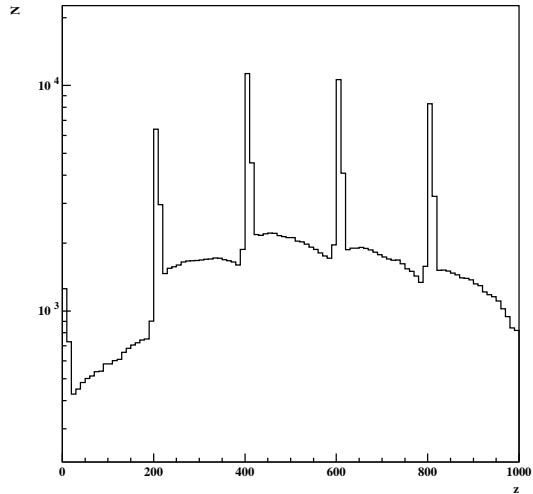} 

\caption{\label{fig2}  
The cascade curve, formed by $10^{12}$ eV iron nuclei in the calorimeter of $5x(1Pb+19Si)$. 
} 
\end{center} 
\end{figure} 

As it can be seen from Figure 2 cascade curve has the piecewise-ordered structure with significant violations associated with huge peaks, which demonstrates the growth of the number of particles in the lead absorber. 

\section{ Mixed calorimeter }

The best option is a mixed calorimeter, which has a target block, leading to the rapid development of the cascade, and homogeneous measuring unit. 
The cascade development process in such a calorimeter $1x(1Pb+99Si)$, consisting of 1 cm of lead and 99 cm silicon is presented in Figure 3. 

\begin{figure}[tbh] 
\begin{center} 
\includegraphics*[width=0.6\textwidth,angle=0,clip]{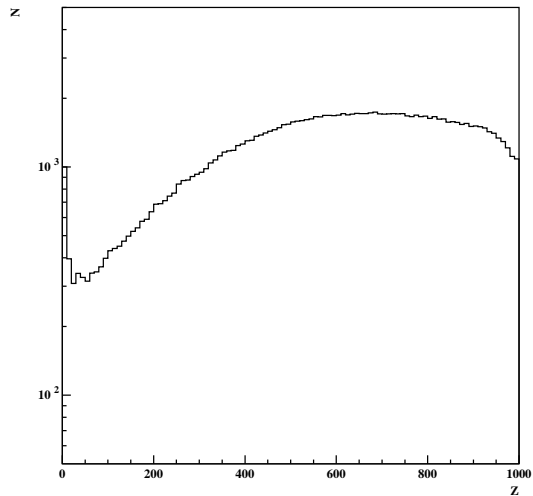} 
\caption{\label{fig3}  
The cascade curve, formed by $10^{12}$ eV iron nuclei in the calorimeter of $1x(1Pb+99Si)$. 
} 
\end{center} 
\end{figure} 

As it can be seen from Figure 3 the cascade curve represents for mixed calorimeter rather ordered structure with a sharp peak in the region of the lead absorber.

For the calorimeter $1x(1Pb+99Si)$ the test measurement of the energy of the simulated cascades, is carried out. For this purpose, 
we performed a simulation of the development of the cascade within the proposed thin calorimeter on the base of software package GEANT4. 

\section{Cascade and correlation curves}

Figure 4 presents the average cascade curves for proton and Fe cascades at energies of $10^{12}$ and $10^{13}$ eV.

\begin{figure}[tbh] 
\begin{center} 
\includegraphics*[width=0.6\textwidth,angle=0,clip]{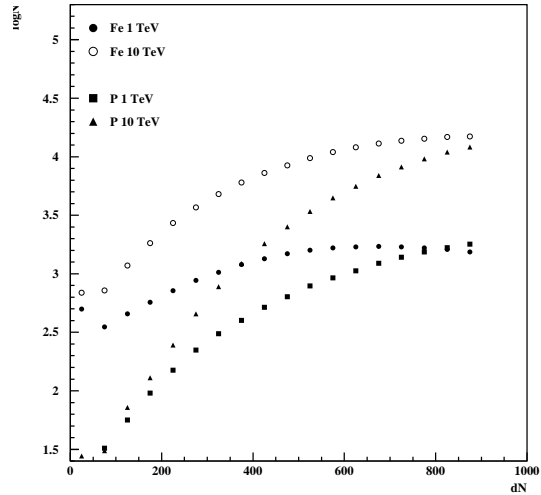} 
\caption{\label{fig4}  
The average cascade curves for proton and Fe cascades at energies of $10^{12}$ and $10^{13}$ eV in the calorimeter of $1x(1Pb+99Si)$.
} 
\end{center} 
\end{figure} 

From figure 4 it follows that the development of proton and Fe cascades is significantly different. 

First, penetration depth of the first interaction is significantly different. 

The most significant factor, determining the observed difference, is due to the value of the cross section. Since the average free path length before interaction is inversely proportional to the cross section, the average depth of the primary interaction for iron is substantially less than for protons.

Figure 5 presents the average correlation curves, i.e. dependence of  the number of particles at a certain level observations on the difference of the number of particles at two levels of observation, separated by a layer of absorber, for proton and Fe cascades at energies of $10^{12}$ and $10^{13}$ eV. The thickness of the layer between measurements of dN is 10 cm, i.e. about 1 radiation units.

\begin{figure}[tbh] 
\begin{center} 
\includegraphics*[width=0.6\textwidth,angle=0,clip]{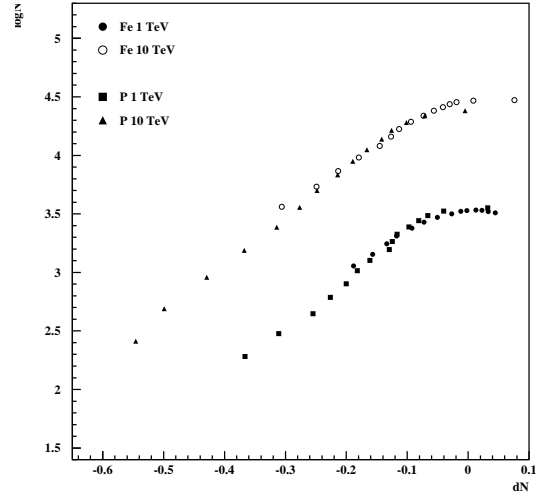} 
\caption{\label{fig5}  
The average correlation curves for proton and Fe cascades at energies of $10^{12}$ and $10^{13}$ eV in the calorimeter of $1x(1Pb+99Si)$.
} 
\end{center} 
\end{figure} 

As it can be seen from Figure 5 the correlation curves represent the ordered structure depending on the primary energy and almost regardless on the type of the primary nucleus.

The size of the detector layer of the calorimeter can be reduced to two radiation units, i.e. up to 19cm. First radiation unit after the lead absorber will be used for the relaxation process, i.e. for the absorption of the particles. The second radiation unit is directly the absorber for the calculation of $dN$.

\section{Test reconstrution of the primary energy}

The analysis procedure consisted of several basic stages:

1. The simulation of cascades, formed by primary protons and iron nuclei with fixed energies.

2. Fitting of  $logN$-$dN$-distributions by polynomial functions of the third order for each fixed energy as function of $logN$ versus $dN$ 
in the form: 

\begin{equation}
log N (dN) =  a_0+a_1dN+a_2dN^2+a_3dN^3                                             
\end{equation}

3. Fitting of coefficients of $a_0$, $a_1$, $a_2$, $a_3$ in dependence on the energy and the construction of the dependence $log N$ 
on $dN$ and $E_0$ in the form:

\begin{equation}
log N (dN,E_0) =  a_0(E_0)+a_1(E_0)dN+a_2(E_0)dN^2+a_3(E_0)dN^3                 
\end{equation}

4. The simulation of test cascades, formed by primary protons and iron nuclei with arbitrary energies, and reconstruction 
of the primary energy on the base of the dependence of (2).

To determine the energy of $i$-th test cascade, using a set of curves (2), it is necessary to vary the $dN$ and $E_0$ so 
that at the same time to minimize the following differences  

\begin{equation}
| log N_{m} - log N (dN,E_0) | < \epsilon 
\end{equation}

\begin{equation}
| dN_{m} - dN | < \epsilon                                                  
\end{equation}

where $logN_{m}$ and $dN_{m}$ are "measured" characteristics of the test cascade at the level of observation. 

By the definition of the energy of test cascades it was obtained that uncertainties of energy reconstruction, $<|(E_i-E_0)/E_0|>$, at the 
estimation of energy by the formulas (1-4), is less than 10 percent.

The accuracy of the energy determination by the standard technique of cascade curves, in the case of registration of the cascade 
to the depth of its maximum development, is about ~50 percent \cite{11}. 

This is clearly argues in favor of the proposed methodology for determining the primary energy, as it significantly reduces the required thickness of the calorimeter and improves the accuracy of the measurements.

\vspace{12pt} 

{\Large \bf Acknowledgements } 

This work was supported by grant N1276/GF2 of Ministry of Education and Science of Kazakhstan Republic.   


\end{document}